\documentclass[aps,prl,twocolumn,superscriptaddress,showpacs]{revtex4}
\usepackage{graphicx,color}

\begin{document}

\title{Unconventional Magnetism in a Nitrogen-Containing Analogue of Cupric Oxide}
\author{A. Zorko}
\affiliation{Jo\v{z}ef Stefan Institute, Jamova 39, 1000 Ljubljana, Slovenia}
\affiliation{EN--FIST Centre of Excellence, Dunajska 156, SI-1000 Ljubljana,
Slovenia}
\author{P. Jegli\v{c}}
\affiliation{Jo\v{z}ef Stefan Institute, Jamova 39, 1000 Ljubljana, Slovenia}
\affiliation{EN--FIST Centre of Excellence, Dunajska 156, SI-1000 Ljubljana,
Slovenia}
\author{A. Poto\v{c}nik}
\affiliation{Jo\v{z}ef Stefan Institute, Jamova 39, 1000 Ljubljana, Slovenia}
\author{D. Ar\v{c}on}
\affiliation{Jo\v{z}ef Stefan Institute, Jamova 39, 1000 Ljubljana, Slovenia}
\affiliation{Faculty of Mathematics and Physics, University of Ljubljana, Jadranska 19,
1000 Ljubljana, Slovenia}
\author{A. Bal\v{c}ytis}
\affiliation{Institute of Applied Research, Vilnius University, Vilnius, Lithuania}
\author{Z. Jagli\v{c}i\'{c}}
\affiliation{Institute of Mathematics, Physics and Mechanics, Jadranska 19, 1000 Ljubljana, Slovenia}
\affiliation{Faculty of Civil and Geodetic Engineering, University of Ljubljana, Jamova 2, 1000 Ljubljana, Slovenia}
\author{X. Liu}
\affiliation{Institute of Inorganic Chemistry, RWTH Aachen University, Landoltweg 1,
D-52056 Aachen, Germany}
\author{A. L. Tchougr\'eeff}
\affiliation{Institute of Inorganic Chemistry, RWTH Aachen University, Landoltweg 1,
D-52056 Aachen, Germany}
\affiliation{Poncelet Lab., Independent University of Moscow, Moscow Center for
Continuous Mathematical Education, Moscow, Russia}
\author{R. Dronskowski}
\affiliation{Institute of Inorganic Chemistry, RWTH Aachen University, Landoltweg 1,
D-52056 Aachen, Germany}
\date{\today }

\begin{abstract}
We have investigated the magnetic properties of CuNCN, the first nitrogen-based analogue of cupric oxide, CuO. Our muon spin relaxation, nuclear magnetic resonance and electron spin resonance studies reveal that classical magnetic ordering is absent down to lowest temperatures. However, large enhancement of spin correlations and unexpected inhomogeneous magnetism have been observed below 80~K. We attribute this to a peculiar fragility of the electronic state against weak perturbations due to geometrical frustration, which selects between competing spin-liquid and more conventional frozen states.
\end{abstract}

\pacs{75.10.Kt, 76.75.+i, 76.60.-k, 76.30.-v}
\maketitle

Numerous copper oxides fascinate with a plethora of exotic electronic ground
states, including high-temperature superconductivity and quantum magnetism
\cite{Lee_2006,Balents_2010}. 
Many of them are low-dimensional spin-1/2 geometrically frustrated antiferromagnets (AFM) \cite{LMM}. There, quantum fluctuations compete with conventional long-range ordering (LRO) and can prevail down to zero temperature, leading to cooperative quantum ground states with spin-liquid (SL) character. In copper oxides, a SL ground state has recently been experimentally reported on two-dimensional (2D) kagom\'e and triangular lattices \cite{Mendels,Zhou}. In the latter case, numerous competing phases -- various SL states, (in)commensurate LRO -- have been theoretically predicted for isosceles triangles, depending on the ratio of the two exchange constants \cite{Weihong,Yunoki_Sorella_2006,Starykh}.

\begin{figure}[b]
\includegraphics[width=1\linewidth, trim=0 3 0 5, clip=true]{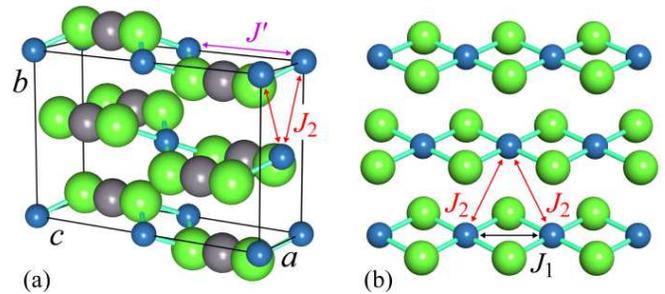}
\caption{(color online). (a) Unit cell of CuNCN with blue Cu,
green N and gray C atoms. Exchange couplings are denoted by $J_1$, $J_2$ and
$J^{\prime }$. (b) Triangular arrangement of Cu atoms in the $ab$ plane.}
\label{fig1}
\end{figure}

Replacing oxygen with other anions is an important step in tailoring their
 physical properties. Transition-metal carbodiimides, $M$(NCN), are good examples for an ``organic'' route. They are
analogues of oxides because the complex anion 
NCN$^{2-}$ is isolobal with O$^{2-}$ \cite{Launay}. Their existence was
elusive until very recently, when metathesis reactions lead to the entire
family, $M$ = Mn, Fe, Co, Ni, and Cu \cite{Liu_2005,Krott_2007}. The
crystal structure of CuNCN (Fig.~\ref{fig1}) is orthorhombic ($Cmcm$) and
consists of corrugated layers with a 4+2 nitrogen environment of Cu$^{2+}$ \cite{Liu_2008,Xiang_2009}. Unlike other $M$(NCN)
materials, all being AFMs
with a N\'eel temperature around 200--300 K, CuNCN exhibits a nearly temperature 
($T$) independent and surprisingly small ($\chi_b \approx 9\cdot 10^{-5}$~emu/mol)
bulk magnetic susceptibility, which enhances at low $T$,
presumably due to a tiny impurity amount \cite{Liu_2008}.

The origin of the surprising reduction in $\chi_b$ has been debated. Based on LDA
calculations, Tsirlin and Rosner predicted a quasi-one-dimensional magnetism, with the leading AFM interaction 
$J^{\prime }\approx 2500$~K bridged via NCN [see Fig.~\ref{fig1}(a)] and LRO around 100~K, due to a ferromagnetic interchain interaction $J_1$ \cite{Tsirlin_2010}. 
Neither bulk magnetic nor neutron diffraction (including polarized
\cite{Xiang_2009}) measurements, however, could confirm LRO. 
Liu \emph{et al.}~\cite{Liu_2008} had proposed an alternative model, 
treating CuNCN as the 2D spatially anisotropic triangular lattice in the \emph{ab} plane [Fig.~\ref{fig1}(b)]. Their GGA+$U$ calculations again predicted strong AFM
exchange couplings ($J_1 \approx$ 800--1000 K) mediated by N atoms. Their size
and the frustrated geometry then lead Tchougr\'eeff and Dronskowski
to propose a 2D resonating-valence-bond (RVB) picture with a nonmagnetic
SL ground state \cite{Tchougreeff_2010}. 
These ambiguities about the ground state call for detailed \emph{microscopic} investigations, allowing detection of intrinsic \emph{spin-only} susceptibility.
In this Letter we present our local-probe study -- including muon spin relaxation 
($\mu$SR), nuclear magnetic resonance (NMR) and electron spin resonance (ESR) --
in the temperature range between 63~mK and 300~K. Around 80 K an unconventional magnetically disordered phase develops implying extreme sensitivity of the magnetic state towards external perturbation. Our results bring forward both similarities and differences between the copper carbodiimide and oxides.

\begin{figure}[t]
\includegraphics[width=1\linewidth, trim=0 20 0 10, clip=true]{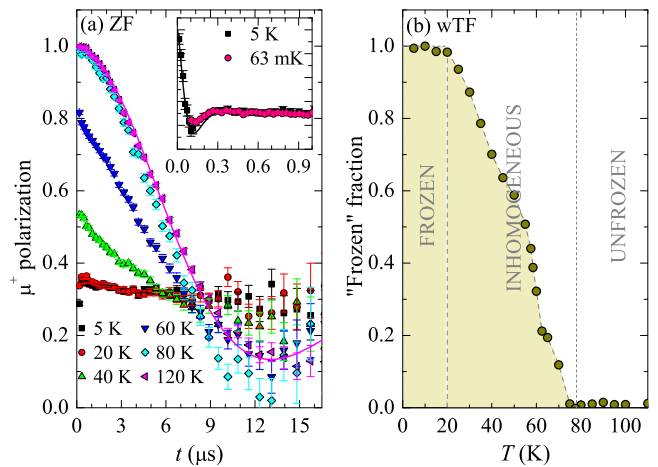}
\caption{(color online). (a) Relaxation of muon spin polarization in zero
magnetic field. Solid lines represent fits to Eq.~(\protect\ref{VKT}), see
text. (b) $T$-dependence of the ``frozen'' fraction, deduced from
weak-transverse-field measurements.}
\label{fig2}
\end{figure}

We first turn to $\mu$SR, which is an extremely sensitive probe for internal magnetic fields. It can easily distinguish between fluctuating and static magnetism on one hand, as well as between LRO and
static disorder on the other \cite{Lee}.
In Fig.~\ref{fig2} we show $\mu$SR results in zero magnetic
field (ZF) and in weak transverse field (wTF) of 2~mT, measured on high-purity powder
samples \cite{foot3} at the MUSR
facility (ISIS, Rutherford Appleton Laboratory, UK) and at the GPS facility
(Paul Scherrer Institute, Switzerland). 
Above 80~K, the ZF muon-spin polarization relaxes according to the relaxation function 
$G(t)=G_{\rm VKT}\cdot \mathrm{exp}[-\lambda t]$ in the Gaussian ($\beta=2$) limit, with the Voigtian Kubo-Toyabe (VKT) function \cite{Crook}
 \begin{equation}
G_{\rm VKT}(t)=\frac{1}{3}+\frac{2}{3} \left[ 1-(\Delta t)^\beta \right]\mathrm{%
exp} \left[ -(\Delta t)^\beta /\beta\right]. \label{VKT}
\end{equation}
\noindent The dominant relaxation is due to static nuclear-magnetic-fields distribution with the width $\Delta/\gamma_\mu = 0.16$~mT ($\gamma_{\mu}=2\pi\cdot 135.5$~MHz/T), while $\lambda = 0.007~\mu$s$^{-1}$ represents a weak dynamical relaxation due to fast
fluctuations of electronic magnetic fields.

Below 80~K the shape of the ZF relaxation function
becomes irregular. The amplitude of the high-temperature component decreases 
at the expense of another, quickly relaxing component [Fig.~\ref{fig2}(a)]. 
The latter prevails below 20~K where the data again fit well to the VKT relaxation function [Eq.~(\ref{VKT})], this time with $\beta=1.44$ and significantly increased static
internal fields, $\Delta/\gamma_\mu=17.7$~mT [inset to Fig.~\ref{fig2}(a)]. The largely
enhanced $\Delta$ can only be attributed to electronic fields. The
observed dip and the ZF polarization leveling at 1/3 for longer times 
\cite{foot} affirm that local fields are \textit{disordered} and \textit{fully frozen} on the muon time scale
\cite{Lee}. We stress that absence of coherent oscillation of polarization discards the
possibility of a LRO state down to at least 63~mK [inset to Fig.~\ref{fig2}(a)]. 

The wTF measurements further disclose the
magnetic state of CuNCN between 80~K and 20~K, where multicomponent ZF
relaxation is observed. The amplitude of the oscillating polarization,
proportional to the fraction of muons still experiencing dynamical electronic
fields, gradually decreases in this range. This witnesses progressive
development of static internal magnetic fields of electronic origin
[``frozen'' fraction in Fig.~\ref{fig2}(b)], whose magnitudes are large
compared to the applied field. This phase is inhomogeneous as part of
the muons detects static and the rest dynamical local fields. The wTF
measurements confirm a fully static phase below 20~K.

The single-dip ZF muon relaxation observed in CuNCN below 20~K is
commonly encountered in spin glasses (SG) \cite{Lee}. SG states have been
detected by $\mu$SR in several frustrated AFMs 
\cite{Dunsinger,Harrison,Wiebe} and appear due to
 some sort of randomness; structural disorder, random fields, bonds or strains,
topological defects, etc. One could therefore hastily assign the CuNCN ground state as a SG state. Such state is generally characterized with a pronounced zero-field-cooled/field-cooled
(ZFC/FC) irreversibility in $\chi_b$ below the freezing
temperature. We have, therefore, searched for the ZFC/FC irreversibility \cite{foot3} and for signs 
of a SG transition in the ac susceptibility in various applied field between 0.8~mT and 5~T, but could not find
any. Another peculiarity of CuNCN, uncharacteristic for SGs, is the anomalous stability of the
inhomogeneous region of coexisting static (low-$T$) and dynamic (high-$T$) phase 
[Fig.~\ref{fig2}(b)]. 

Although low-$T$ $\mu$SR results in CuNCN reveal static frozen fields, it is the absence of thermal remanence and well defined freezing temperature that rules out the canonical SG scenario. 
A similarly inhomogeneous phase has been reported in
the frustrated kagom\'e compound vesignieite and reasoned to be due
to a vicinity of a quantum critical point \cite{Colman}. On the other hand, in few spin-gap systems partially frozen disordered states were unexpectedly detected by $\mu$SR \cite{Andreica,Fudamoto}. They were attributed
to a muon locally perturbing exchange pathways thus prohibiting singlet formation and liberating spin degrees
of freedom in its vicinity \cite{Chakhalin,Aczel}. For slow dynamics, a local-probe muon
would not distinguish such a state from collective spin freezing. This is
plausible in CuNCN, as muons are likely to stop near bridging N atoms
of the NCN$^{2-}$ groups, similarly as protons are positioned in related $M$%
(NCNH)$_2$ ($M$ = Fe--Ni) \cite{hydrocyanamide}.

To avoid any disturbance of the CuNCN structure by external probes we next decided to employ
$^{14}$N ($I=1$) NMR measurements because intrinsic $^{14}$N nuclei also directly detect
local magnetic fields. To obtain a reasonable signal-to-noise ratio the frequency-swept spectra were measured in high magnetic field of 9.4~T \cite{foot4}. 
\begin{figure}[t]
\includegraphics[width=1\linewidth, trim=0 8 0 19, clip=true]{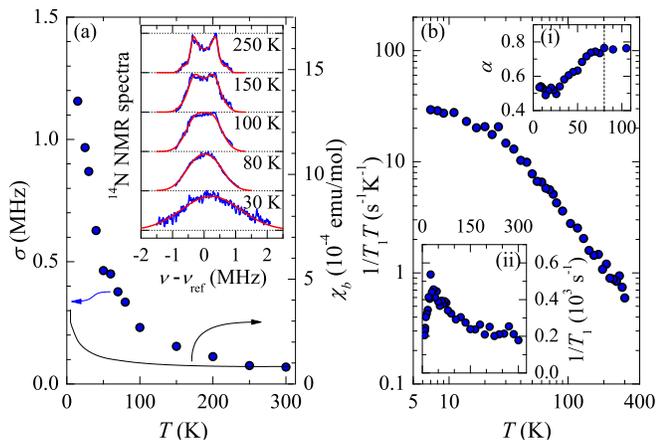}
\caption{(color online). (a) Comparison of Gaussian
broadening $\sigma$ of $^{14}$N NMR spectra and bulk susceptibility $\protect\chi_b$ (in 5~T). Inset: Simulation (red) of selected experimental (blue) spectra. (b) Spin-lattice relaxation rate 
$1/T_1$ divided by $T$. Inset: $T$-dependence of (i) the
stretch exponent and (ii) $1/T_1$.}
\label{fig3}
\end{figure}
At 250~K the NMR spectrum (Fig.~\ref{fig3}) exhibits a characteristic
quadrupolar powder line shape, which allows fixing the quadrupolar
interaction parameters, $\nu_Q = 0.79(2)$~MHz and $\eta_Q = 0.06(2)$.
In order to simulate the low-$T$ spectra we assume a convolution with a $T$-dependent
Gaussian local field distribution with the width $\sigma(T)$ originating from the hyperfine coupling between the nuclear and Cu$^{2+}$electronic moments. $\sigma$ starts increasing dramatically below ca.\ 100~K [Fig.~\ref{fig3}(a)] but fails to
scale with the moderate increase of $\chi_b$ at low $T$ [Fig.~\ref{fig3}(a)], thus ruling out the distribution of hyperfine couplings as the source of the spectral broadening.
Moreover, spin-spin relaxation measurements (not shown) discard an increase in the relaxation rate
as a possible origin of the low-$T$ broadening, which is therefore inhomogeneous and due to
a broad distribution of local susceptibilities ({\it static electronic moments}) at low $T$ and high fields.

The $T$ dependence of the $^{14}$N spin-lattice relaxation rate $1/T_1$, measured with an inversion recovery technique,
reveals the development of electronic spin dynamics since the fast
spin-lattice relaxation times of only a few ms [Fig.~\ref{fig3}(b)] directly
measure the local-field fluctuations weight at the Larmor frequency. The 
$^{14}$N magnetization relaxation curves were analyzed with the stretched
exponential model $M_z(t)-M_0 \propto \exp[-(t/T_1)^{\alpha}]$. Although 
$1/T_1$ shows no obvious anomaly at 80~K, the stretch exponent $\alpha$
clearly starts to decrease below 80~K and reaches a value of 0.5 at lowest 
$T$ [inset (i) to Fig.~\ref{fig3}(b)]. This evidences that the distribution of
relaxation rates widens below 80~K --- another hallmark of a magnetically
highly disordered state. Around 20~K a $1/T_1$ maximum is observed [inset
(ii) to Fig.~\ref{fig3}(b)], indicating freezing of spin dynamics below this
temperature. The freezing is moderate, far from opening of a spin gap which
would require a suppression of $1/T_1T$, not observed experimentally 
[Fig.~\ref{fig3}(b)]. A ZF spin gap could however be closed by the large applied field of 9.4~T \cite{Starykh}.
\begin{figure}[t]
\includegraphics[width=1\linewidth, trim=0 5 0 15, clip=true]{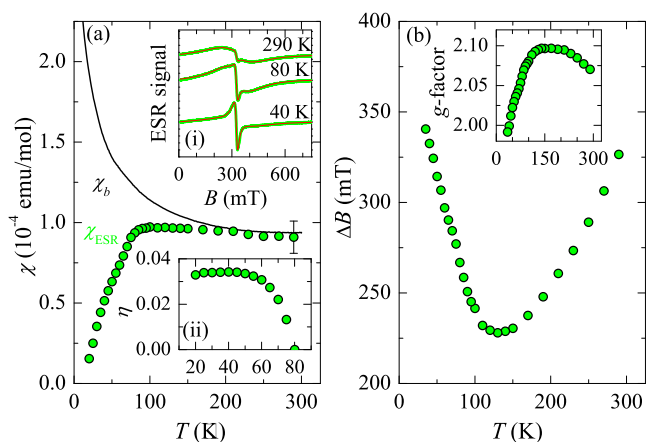}
\caption{(color online). (a) Comparison of bulk susceptibility $\protect\chi_b$ and ESR spin-only susceptibility $\protect\chi_{\rm ESR}$, both measured in 0.33~T. Inset: (i) Fits (red) of selected ESR spectra (green) and (ii) $T$-dependence of the 2D RVB order parameter. (b) $T$-dependence of the ESR line-width and the $g$-factor (inset).}
\label{fig4}
\end{figure}

The electronic properties of CuNCN were, therefore, also investigated at much lower fields ($B_0\approx 0.33$~T) with continuous-wave ESR measurements in X-band ($\nu_L = 9.6$~GHz), following both FC and ZFC temperature treatments. The ESR spectra 
[inset (i) to Fig.~\ref{fig4}(a)] 
comprise two overlapping components; a broad Lorentzian component with the
RT line-width of 280(10)~mT and a ca.~10-times
narrower Gaussian component. The RT $g$-factor value $g=2.07$ of the broad component
is common for Cu$^{2+}$ in a square planar
environment \cite{AbragamBleaney}. The broad-component ESR intensity, corresponding to the spin-only susceptibility,
was calibrated at RT with a couple of standards, $\chi_{\rm ESR}^{\rm RT} \approx
9(1)\cdot 10^{-5}$~emu/mol.
It is the same as the bulk susceptibility, which proves that the broad component represents the intrinsic ESR signal. The
narrow component, on the other hand, is ca.\ 100-times less intense and exhibits a Curie-like $T$-dependence. The total fraction of impurity spins contributing to it is estimated as
only $0.04$\%. 
The intrinsic ESR component exhibits a minimum in the line-width 
and a maximum in $g$-factor around 130~K [Fig.~\ref{fig4}(b)]. Both indicate growth of spin correlations 
\cite{AbragamBleaney}, which appear particularly strong below $\sim 80$~K. The featureless $T$-dependence of both parameters and no obvious ZFC/FC dissimilarities of ESR spectra speak against a conventional SG transition, in line with the lack of usual susceptibility fingerprints of such a state.
Moreover, the intrinsic $\chi_{\rm ESR}$ rapidly diminishes below 80~K, thus diverging from $\chi_b$ [Fig.~\ref{fig4}(a)]. 

The suppressed $\chi_{\rm ESR}$ implies opening of a spin gap and thus allows for an alternative interpretation of the magnetic state in CuNCN. For the anisotropic triangular lattice, Hayashi and Ogata proposed that in ZF two spin-singlet states
are formed at different temperatures, depending on the $J_2/J_1$ ratio
\cite{Hayashi_Ogata}. Below the upper critical temperature $\theta_c=3J_1/8$
a quasi-1D gapless RVB phase sets in, while a 2D gapped RVB
phase appears at the lower critical temperature
$\theta^{\prime }_c=J_2/8(1-2J_2/3J_1)$ \cite{Tchougreeff_2010}. 
We have further developed the analytical
susceptibility expression \cite{Tchougreeff_2010} for the 2D RVB phase and
arrive at
\begin{equation}
\chi=\mu_{B}^{2}\left\{\frac{(1-ca)f(3\sqrt{2}J_2\eta)}{3\sqrt{2}%
\left(J_1\xi-J_2\eta\right)}- ca\frac{\partial
f}{\partial\varepsilon}(3\sqrt{2}J_2\eta)\right\}, \label{EqChi2}
\end{equation}
\noindent where $a=J_2\eta /J_1\xi$, with $\xi$, $\eta$ as the 1D, 
2D RVB order parameters, respectively; $f$ is a Fermi distribution
function and $c = 1 + (1 - \frac{3B}{4}+\frac12 B\ln a)\frac{a}{1-a}$
\cite{foot2}. Assuming constant $\xi=1/\sqrt{2}\pi$ \cite{Tchougreeff_2010} in the 2D RVB phase we estimate the exchange constants $J_1=2300$~K and $J_2=540$~K and establish the $T$-dependence of the 2D RVB order
parameter [inset (ii) to Fig.~\ref{fig4}(a)] from $\chi_{\rm ESR}(T)$ for $\theta^{\prime }_c=80$~K. 

We note that both RVB phases are characterized by complete
spin-pairing, in agreement with no magnetic neutron scattering in polarized
experiments \cite{Xiang_2009}.
The growing spin correlations observed in ESR are also in line with the
2D RVB phase below $\theta^{\prime }_c$. This phase, however, appears to be very fragile, as muons can locally distort singlets and lead to static local states within the spin gap $\Delta_s$ below $\theta^{\prime }_c$. Moreover, as $\Delta_s= 3\sqrt{2}J_2\eta$ is comparable to the applied magnetic field in NMR, it is reasonable to assume that the lack of the spin gap and the growing static fields observed in NMR may be field induced. This calls to mind the recently discovered emergence of inhomogeneous moments in extremely weak fields on another triangular lattice \cite{Pratt} and points to the importance of frustration which amplifies the role of small perturbations in selecting between competing spin-liquid and more conventional frozen phases.

In conclusion, CuNCN, the first nitrogen-based analogue of CuO, with a
surprisingly low magnetic susceptibility, lacks magnetic
ordering down to at least 63~mK. A magnetically highly disordered state has been witnessed below 80~K by complementary local probes. Alternatively to a SG-like frozen state featuring an unprecedented broadness of inhomogeneous static and dynamic phase, no well-defined freezing temperature and no conventional susceptibility-fingerprint SG features, a dynamical spin liquid is proposed. The fragility of the electronic state to perturbations seems to be a peculiarity of carbodiimides as opposed to more conventional oxides. 

P.~M\"uller, V.~Feldman, F.~Haarmann, P.~Mendels and F.~Bert are gratefully acknowledged for
valuable discussion. We thank H. Luetkens and A. D. Hillier for technical
assistance with $\mu$SR. We
acknowledge the financial support of the Slovenian Research Agency (project
J1-2118), the Russian Foundation for Basic Research (grant no.~10-03-00155)
and the German Science Foundation. The $\mu$SR measurements were supported
by the European Commission (contract CP-CSA\_INFRA-2008-1.1.1 Number
226507-NMI3).

\begin{widetext}
\begin{center}
\large{\bf Supplementary information:

Unconventional Magnetism in a Nitrogen-Containing Analogue of Cupric Oxide}\\
\vspace{0.5cm}
\small{A. Zorko, P. Jegli\v{c}, A. Poto\v{c}nik, D. Ar\v{c}on, A. Bal\v{c}ytis, Z. Jagli\v{c}i\'{c}, X. Liu, A. L. Tchougr\'eeff, and R. Dronskowski}
\end{center}
\vspace{0.5cm}
\end{widetext}

\section {Synthesis and Crystal structure}

\begin{figure}[h]
\includegraphics[width=0.92\linewidth, trim=10 5 0 5, clip=true]{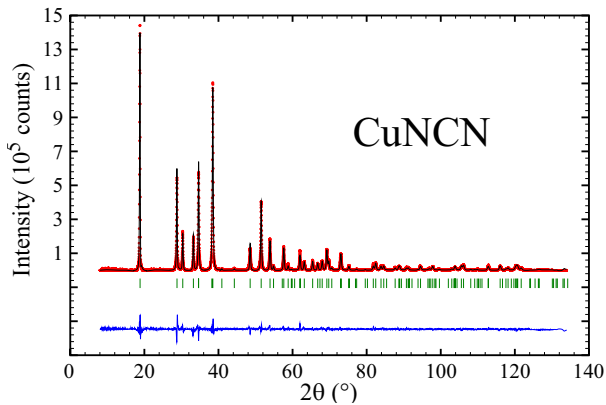}
\caption{(color online). Observed (red), calculated (black), and difference (blue) intensities of the X-ray Rietveld refinement of single-phase CuNCN using strictly monochromatized Cu-K$\alpha_1$ radiation ($\lambda$ = 1.54056 \AA) with a flat-sample geometry and an imaging-plate detector. The vertical positions of the Bragg reflections are given in green. The residual values are 0.038 ($R_{\rm p}$), 0.052 ($R_{\rm wp}$) and 0.035 ($R_{\rm Bragg}$).}
\label{fig1s}
\end{figure}

All preparative steps were based on high purity educts (99.9999\% Cu, 99.999\% HCl, highly crystalline cyanamide, milli-Q water) or on educts of analytical grade (concentrated NH$_3$ solution, Na$_2$SO$_3$). After having synthesized the precursor Cu$_4$(NCN)$_2$NH$_3$,[1] it was oxidized over night in aqueous solution to yield X-ray pure CuNCN of excellent crystallinity [2], with noticeable Bragg peaks even at highest angles (Fig.~\ref{fig1s}). Neutron diffraction data [3] did not show any indication of defect formation on any of the atomic sites.

\section {Bulk magnetic susceptibility}

\begin{figure}[h]
\includegraphics[width=0.97\linewidth, trim=0 10 0 15, clip=true]{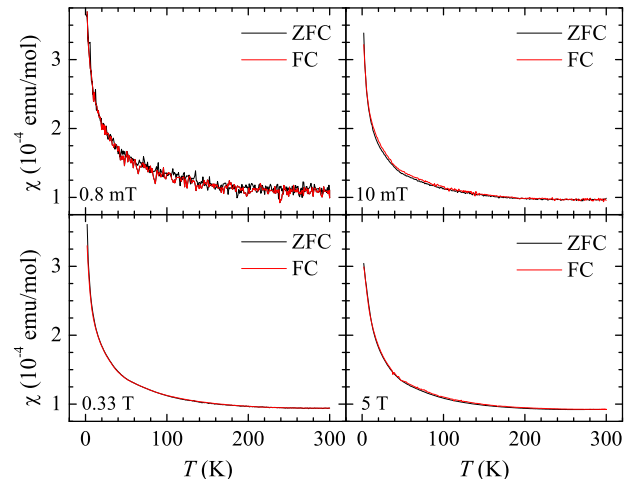}
\caption{(color online). $T$-dependence of ZFC and FC bulk susceptibility in various applied magnetic fields.}
\label{fig2s}
\end{figure}

Magnetic susceptibility ($\chi=M/B$) was measured on a 90~mg powder sample between 2~K and 300~K with a Quantum Design SQUID magnetometer. Zero-field-cooled (ZFC) measurements were performed after cooling the sample to the base temperature in zero field and field-cooled (FC) measurements after cooling in applied magnetic field. In Fig.~\ref{fig2s} we show measurements in various applied magnetic fields $B$ in the range between 0.8~mT and 5~T. There is no obvious ZFC/FC splitting emerging at low temperatures that would be characteristic of a canonical spin glass transition in any of the applied fields. 

\small{
\noindent $[1]$ X. Liu \textit{et al.}, Z. Anorg. Allg. Chem. \textbf{631}, 1071 (2005).\\
$[2]$ X. Liu \textit{et al.}, Z. Naturforsch. B \textbf{60}, 593 (2005).\\
$[3]$ X. Liu \textit{et al.}, J. Phys. Chem. C \textbf{112}, 11013 (2008).}


\begin{thebibliography}{99}
\bibitem{foot} In powders polarization of 1/3 of all muons does not relax
because it is parallel to internal field.
\bibitem{foot4} We used pulse sequence 
${\beta}-{\tau}-{\beta}-{\tau}-\mathrm{echo}$, pulse length $\tau_{\beta}=8~\mathrm{\mu}$s, interpulse
delay $\tau = 60~\mathrm{\mu}$s, and repetition time 100~ms at RT. $^{14}\mathrm{N}$ reference frequency of 28.9027~MHz was determined
from the CH$_3$NO$_2$ standard.
\end{thebibliography}
\end{document}